# Effective Beam Width of Directional Antennas in Wireless Ad Hoc Networks

Jialiang Zhang and Soung-Chang Liew

*Abstract* —It is known at a qualitative level that directional antennas can be used to boost the capacity of wireless ad hoc networks. Lacking is a measure to quantify this advantage and to compare directional antennas of different footprint patterns. This paper introduces the concept of the *effective beam width* (and the *effective null width* as its dual counterpart) as a measure which quantitatively captures the capacity-boosting capability of directional antennas. Beam width is commonly defined to be the directional angle spread within which the main-lobe beam power is above a certain threshold. In contrast, our *effective beam width* definition lumps the effects of the (i) antenna pattern, (ii) active-node distribution, and (iii) channel characteristics, on network capacity into a single quantitative measure. We investigate the mathematical properties of the effective beam width and show how the convenience afforded by these properties can be used to analyze the effectiveness of complex directional antenna patterns in boosting network capacity, with fading and multi-user interference taken into account. In particular, we derive the extent to which network capacity can be scaled with the use of phased array antennas. We show that a phased array antenna with *N* elements can boost transport capacity of an Aloha-like network by a factor of order $N^{1.620}$.

Manuscript received March 31, 2007. This work was supported by the Areas of Excellence scheme (Project Number AoE/E-01/99) and the Competitive Earmarked Research Grant (Project Number 414305) established under the University Grant Committee of the Hong Kong Special Administrative Region, China.

The authors are with the Department of Information Engineering, Chinese University of Hong Kong, ShaTin, N.T., Hong Kong (e-mail: jlzhang@ie.cuhk.edu.hk; soung@ie.cuhk.edu.hk).

## I. INTRODUCTION

Qualitatively and intuitively, it is widely accepted that the use of directional antennas in wireless ad hoc network can reduce mutual interference and improve spatial reuse to boost network capacity. Recent work [1]-[6] showed that this advantage can be related to the antenna "beam width". In particular, attempts have been made to "quantify" the merit of directional antennas using simple antenna characteristics, such as main-lobe beam width [1]-[5], suppression ratio [1], and number of simultaneous beams with infinitesimal beam widths [6]. Investigations at the system level have been largely based on simplistic models of antenna patterns that are unrealizable (e.g., flat-topped antenna model in [1], circular sector model in [3], and infinitesimal-beam-width model in [6]). In addition, the effects of cumulative multi-user interference and channel fading have not been taken into account.

Meanwhile, antenna-theory experts have suggested physically-motivated beam-width definitions, such as antenna directivity, half power beam width (HPBW) and first null beam width (FNBW), as metrics for measuring the "goodness" of general antenna patterns [7]. These physically-motivated definitions, however, have not provided for the context under which the antennas will be used. In wireless ad hoc networks, the coupling of the antenna pattern with the active-node distribution and channel-gain variation must be taken into account to gauge the effect of the antenna pattern on network performance.

Openly missing is a generic measure to quantify the advantage of arbitrary directional antennas under general active-node distributions and channel states. This paper is a first attempt to fill this gap. The main contribution of this paper is the introduction and investigation of the *effective beam width* of general antenna patterns. The effective beam width is a quantitative measure that captures the network-capacity-boosting capability of arbitrary directional antennas. Our definition of the effective beam width is motivated by the observation that "interference cancellation" in a wireless network is due to the combined effect of (i) antenna pattern, (ii) active-node distribution, and (iii) channel characteristics. In a nutshell, the effective beam width is a performance measure of the integrated effect of the three attributes. As will be shown in this paper, this performance measure enjoys a number of convenient mathematical properties that greatly facilitate our analysis of the effectiveness of complex directional antenna patterns for boosting network capacity.

Of particular intellectual interest is a fundamental analytical relationship between two scenarios: (i) cumulative multi-user interference with Rayleigh fading, and (ii) pair-wise interference with no fading. This relationship allows us to apply the concept of effective beam width to both scenarios. We will see how network capacity scales as the effective beam width varies in both cases.

The rest of this paper is organized as follows. Section II reviews multiple-interference and pair-wise-interference models for our later analysis. Section III puts forth the concept of effective beam width (and its counterpart, effective null width) based on the pair-wise-interference model. The properties of the effective beam width are investigated. Section IV studies the impact of effective beam width on the scalability of network capacity. In particular, subsection IV.F extends the result to the scenario where multiple-interference model and Rayleigh fading channel are assumed. Section V concludes the paper.

## II. INTERFERENCE MODELS FOR DIRECTIONAL ANTENNAS

### A. Propagation Model

Consider an *interference-limited* planar wireless network (where the ambient noise is negligible) consisting of a group of $n$ wireless stations (denoted by $\mathcal{N} = \{1, 2, \ldots, n\}$). When a transmitter $T$ transmits at power level $P_T$, the power received by a receiver $R$ is $H_{RT} P_T$, where $H_{RT}$ is the channel gain from $T$ to $R$. The channel gain $H_{RT}$ depends on the aggregate effects of path loss, shadowing, channel fading and antenna gain. In this paper, we adopt the following model [8]:

$$H_{RT} = G_T F_{RT} G_{RT}^{PL} G_R, \qquad (1)$$

where $G_{RT}^{PL}$ is the path loss (or path gain) from $T$ to $R$; $F_{RT}$ models the channel fluctuation; and $G_T$ ($G_R$) is the antenna gain of transmitter $T$ (receiver $R$). We further assume the two-ray propagation model [8] which models the path-loss attenuation as a function of distance:

$$G_{RT}^{PL} = K / |TR|^\alpha, \alpha \geq 1, \qquad (2)$$

where $|TR|$ is the Euclidean distance between $T$ and $R$; $\alpha$ is the path-loss exponent; and $K$ is a normalization factor. In this paper, we will consider two scenarios of channel fluctuation: no fading and Rayleigh fading. For different $\{R, T\}$ pairs, under no fading, $F_{RT}$ are unity constant; under Rayleigh fading, $F_{RT}$ are independent exponentially distributed random variables with unit mean [9].

Antenna gain $G_T$ ($G_R$) is the central focus of this paper. For the omni-directional antenna, antenna gain is constant independent of azimuthal angle. For an arbitrary directional antenna, antenna gain is generally a function of azimuthal angle. Intuitively, when omni-directional antennas are used in the wireless network for unicast communication, the excessive power radiated at unwanted directions causes severe mutual interference among the nodes. This in turn leads to poor spatial reuse and network-capacity degradation [10], [11]. The power of a directional antenna can be directed to a desired focal orientation between the transmitter and receiver of a link to minimize mutual interference with other links. This can potentially lead to better spatial reuse to yield higher network capacity.

For our studies, we refine our definition of antenna gain $G_T$ ($G_R$) as the *normalized* antenna power pattern at $T$ ($R$). Specifically, $G_T$ ($G_R$) is a function of azimuthal angle $\theta$, where $\theta$ is the angle with respect to the direction at which the antenna is pointing. By definition, $\theta = 0$ is the boresight direction;

$G_T(0) = G_R(0) = 1$; and for all $\theta$,

$$0 \leq G_T(\theta) \leq 1, \quad 0 \leq G_R(\theta) \leq 1. \qquad (3)$$

In the special case of the omni-directional antenna, $G_R(\theta) = G_T(\theta) \equiv 1$ for all $\theta$.

Let $\theta_R$ ($\theta_T$) denote the angle between the line connecting the transmitter $T$ and receiver $R$, and the boresight direction of the transmitter (receiver) antenna. From the above discussion, we can establish the following propagation model:

$$P_R = K \cdot F_{RT} \cdot P_T \cdot G_T(\theta_R) \cdot G_R(\theta_T) / |TR|^\alpha, \alpha \geq 1, \qquad (4)$$

where $P_T$ is *redefined* as the transmitted power of $T$ at the antenna boresight; $P_R$ is the received power at $R$.

### B. Physical Link Interference Models of Generic Directional Antenna

We assume at time instant $t$, a set of active links (transmitter-receiver pairs), $\mathcal{S}_{act}(t)$ is selected by certain media access control (MAC) protocol to transmit simultaneously. Let $T_i$ and $R_i$ denote the transmitter and receiver of an active link $i$. (For brevity, we will also use $T_i$ and $R_i$ to denote their positions.) We assume that by proper beam steering, the transmitter and receiver of a link can point their antennas directly at each other so that the power is the highest along the link orientation. Thus, for each link $i$, $G_{R_i}(\theta_{T_i}) = G_{T_i}(\theta_{R_i}) = 1$. We also assume each active transmitter will use equal power $P$ to transmit at its boresight direction.

From (4), the signal power received at $R_i$ (denoted by $S_i$) is given by $S_i = K \cdot F_{R_i T_i} \cdot P / d_i^\alpha$, where $d_i$ is the link length of link $i$; and the interference from link $j$ to link $i$ (denoted by $I_{ij}$) is given by $I_{ij} = K \cdot F_{R_i T_j} \cdot P \cdot G_{R_i}(\theta_{ij}) \cdot G_{T_j}(\phi_{ji}) / |T_j - R_i|^\alpha$, where $\theta_{ij} = \angle T_i R_i T_j, \phi_{ji} = \angle R_j T_j R_i$ (see Fig. 1). Let $SIR_0$ (>1) be the signal-to-interference ratio (SIR) threshold required for proper detection. For reception at link $i$ not to be corrupted by multi-user interference from other active links, we require

$$SIR_i = S_i / \sum\nolimits_{j \in \mathcal{S}_{act}(t), j \neq i} I_{ij} \geq SIR_0. \qquad (5)$$

The above is based on the "multiple-interference model", since the aggregate interferences of other users are considered. In this paper, we will first look at a more tractable "pairwise-interference model", which paves the way for the extension to the multi-interference model later. For the pair-wise interference model, we relax the requirement in (5) to

$$S_i / I_{ij} \geq SIR_0, \forall j \in \mathcal{S}_{act}(t) \text{ and } j \neq i. \qquad (6)$$

Thus, adopting the pair-wise interference model in our analysis will lead to an upper-bound estimation of the network capacity. In the following discussion, we will show how to make use of its convenient geometrical properties. The definition of our *effective beam width* is built upon this model, and generalization to the multiple-interference model will be discussed in Subsection IV.F.

To begin, let us consider the no-fading scenario where $F_{R_i T_j} \equiv 1, \forall i, j \in \mathcal{S}_{act}(t)$. With reference to Fig. 1, define the "guard zone" $\Delta \triangleq SIR_0^{1/\alpha} - 1 > 0$ [10]; and $G^*(\theta) \triangleq G(\theta)^{1/\alpha}$,

$0 \leq G^*(\theta) \leq 1$. Then $S_i / I_{ij} \geq SIR_0$ in (6) can be rewritten as

$$|T_j - R_i| \geq (1+\Delta) d_i \cdot G^*_{R_i}(\theta_{ij}) \cdot G^*_{T_j}(\phi_{ji}). \quad (7)$$

In the omni-directional case, $G_{R_i}(\theta_{ij}) = G_{T_j}(\phi_{ji}) \equiv 1$, $\forall \theta_{ij}, \phi_{ji}$. Hence, our pair-wise-interference model is consistent with and is a generalization of the "protocol model" in [10].

### C. Potential Interference Region

The interference region is a critical geometrical region useful for characterizing spatial reuse and estimating network capacity [10], [11]. Define the *potential interference region* of an active link $i$ as a vulnerable area associated with $R_i$ within which the transmission of $T_j$ ($j \neq i$) *may* interfere with the transmission from $T_i$ to $R_i$. Particularly, the potential interference region of link $i$ is $PI_i = \{x : |x - R_i| \leq (1+\Delta) d_i\}$, according to (3) and (7). The active transmitters (other than $T_i$) within this region are referred to as *potential interfering neighbors* of $R_i$ (link $i$).

### D. Antenna Pattern and Phased Array Antenna

From (7), we note that the antenna pattern affects mutual interference between links. Throughout this paper, unless otherwise specified, we assume an arbitrary directional antenna pattern. An interesting question is as follows: *How to characterize the goodness of an antenna pattern in terms of its interference cancellation capability using a single quantitative measure?* Our answer to this question will be given in the next section.

Since we will also investigate the phased array antenna as a special case in this paper, we provide a brief review of it here. A phased array antenna is an array of antennas whose phases can be adjusted to control the shape of the beam and the directions of the nulls electronically [12]. With $N+1$ antenna elements spaced at distance $D$ ($D \leq \lambda/2$) apart in an linear array, an array factor

$$AF_N(\theta) = |\sum_{k=0}^{N} a_k e^{-2k\pi iD \sin\theta / \lambda}| \quad (8)$$

is achievable [13], where $\lambda$ is the operating wavelength, and $a_k$ ($0 \leq k \leq N$) are arbitrary complex numbers. Specifically, we can arbitrarily position $N$ nulls at orientation $\{\theta_1, \theta_2, ..., \theta_N\}$ with the array factor

$$AF_N(\theta) = |\prod_{s=1}^{N} \{e^{-2\pi iD \sin\theta / \lambda} - e^{-2\pi iD \sin\theta_s / \lambda}\}|, \quad (9)$$

where $a_k = (-1)^{N-k} \sum_{S_{N-k}} [\exp(-2\pi iD \sum_{\theta_s \in S_{N-k}} \sin\theta_s / \lambda)]$, $k < N$ and $a_N = 1$; $S_k$ is a subset consisting of $k$ elements of $\{\theta_1, \theta_2, ..., \theta_N\}$ (there are altogether $\binom{N}{k}$ such subsets), according to Vieta's Formulas. Define $\theta_m = \arg\max_\theta AF_N(\theta)$. (If there are multiple maxima in $AF_N(\theta)$, we can just simply select one of them to be $\theta_m$.) Note that the normalized magnitude of the electric field $G_E(\theta)$ is proportional to the array factor. Therefore we can express the normalized antenna power pattern as

$$G_N(\theta) = G^2_{EN}(\theta) = AF^2_N(\theta + \theta_m) / AF^2_N(\theta_m). \quad (10)$$

A concrete example of normalized antenna power pattern ($N = 4$, $D = \lambda/2$, $\theta_s = 2\pi s / (N+1)$ in (9)) is shown in Fig. 2.

Intuitively, when we use the main beam at $\theta = 0$ to transmit (receive), we wish the power radiated (received) at other angles, especially in which there is a potential interfered receiver (interfering neighbor) to be as weak as possible. A sharp main beam, a good deal of nulls and small side lobes is favorable to this strategy. In the later analysis, we will turn this qualitative argument into a quantitative result.

## III. EFFECTIVE BEAM WIDTH OF DIRECTIONAL ANTENNAS

### A. Concept of Null width and Beam width

We now apply the interference model to define the *effective beam width* (and its dual counterpart, effective null width) of an arbitrary directional antenna pattern. Given any antenna pattern $G(\theta)$, it is clear that the azimuthal distribution of its magnitude affects the performance of interference cancellation. For a general $G(\theta)$ and a threshold $\beta \geq 0$, define the *null width* of $G^*(\theta)$ with respect to threshold $\beta$ as the total directional angle spread in which $G^*(\theta)$ is no greater than $\beta$, namely, $|\{\theta \,|\, G^*(\theta) \leq \beta\}|$, where $|\Omega|$ is the Lebesgue measure of the set $\Omega$. $|\{\theta \,|\, G^*(\theta) \leq \beta\}|/2\pi$ is the corresponding *normalized null width*; whereas the *normalized beam width* is defined as its complement, i.e., $|\{\theta \,|\, G^*(\theta) > \beta\}|/2\pi$. Given a specific $G^*(\theta)$, the null width is an increasing function of threshold $\beta$ (see Fig. 3).

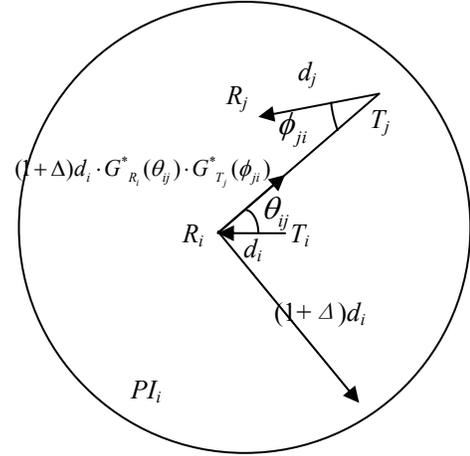

Figure 1.   Interference from node $T_j$ to link $i$

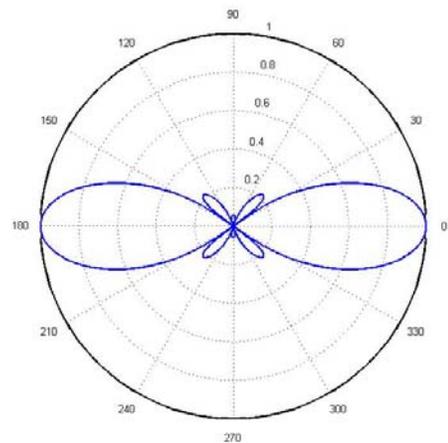

Figure 2.   An example of normalized antenna power pattern (To show the nulls clearly, we plot $G_N^*(\theta) (= G_N^{1/\alpha}(\theta))$ instead of $G_N(\theta)$ in this figure, where $\alpha = 4$. The shape of $G_N(\theta)$ is similar to that of $G_N^*(\theta)$.)

Our definitions of the *effective* null and beam widths to be presented shortly are different from the above normalized null and beam widths, since the former needs to also take into account the effect of active-node distribution on mutual interference.

To clarify the relationship between beam width and interference, consider Fig. 1 again. Denote the event $E_I$ = {*potential interfering neighbor $T_j$ can interfere with link i*}. Consider a potential interfering neighbor at normalized distance $X \triangleq |T_j - R_i|/(1+\Delta)d_i, (X \leq 1)$ from our reference receiver $R_i$. Then, according to (7), event $E_I$ occurs iff

$$G^*_{R_i}(\theta_{ij}) G^*_{T_j}(\phi_{ji}) > X . \quad (11)$$

Hence, we have

$$\Pr(E_I) = \Pr(G^*_{R_i}(\theta_{ij}) G^*_{T_j}(\phi_{ji}) > X) . \quad (12)$$

We start with the simplest case assuming that the potential interfering neighbor $T_j$ is uniformly distributed in the potential interference region $PI_i$, with an independently chosen receiver $R_j$ uniformly distributed within its transmission range $r$. Recall a well-known property of uniform distribution [14] as follows:

Let polar coordinates $(\rho,\theta)$ represent an arbitrary point located in a 2-Dimensional space. The region defined by $\rho \leq R_0$ is a disk of radius $R_0$ centered at the origin. A node's position $(\rho,\theta)$ is uniformly distributed within such a disk if and only if $\rho$ and $\theta$ are independent, and their probability density functions $p(\rho)$ and $p(\theta)$ are:

$$p(\rho) = 2\rho/R_0^2, 0 \leq \rho \leq R_0; \quad p(\theta) = 1/2\pi, 0 \leq \theta < 2\pi . \quad (13)$$

Consequently, $\theta_{ij}$ and $\phi_{ji}$ are i.i.d. uniformly distributed within $[0, 2\pi)$, whereas the probability density function of $X$ is $p_X(x) = 2x, 0 \leq x \leq 1$ (and the corresponding probability distribution function is $F_X(x) = \Pr(X \leq x) = x^2, 0 \leq x \leq 1$). All of them ($\theta_{ij}$, $\phi_{ji}$ and $X$) are mutually independent. For convenience, we define random variables $Y$ and $Z$ as follows:

$$Y = G^*_{R_i}(\theta_{ij}), \quad Z = G^*_{T_j}(\phi_{ji}) ;$$

and their corresponding probability distribution function as $F_Y(y)$ and $F_Z(z)$. Random variables $Y$, $Z$ and $X$ are also mutually independent. Eq. (12) can be rewritten as

$$\Pr(E_I) = \Pr(YZ > X) . \quad (14)$$

Note that in our "simplest case", the distributions of $X$, $Y$ and $Z$ are independent of $(1+\Delta)d_i$, the radius of the potential interference region $PI_i$, thanks to the nice property of uniform distribution. Hence, the link length $d_i$ and guard zone $\Delta$ will not affect $\Pr(E_I)$.

*Directional Transmission and Omni Reception:*

To reveal the relationship between beam width and interference, we first deal with the scenario where transmission uses directional antennas and reception uses omni-directional antennas. We have $Y \equiv 1$ for all $\theta_{ij}$ within $[0, 2\pi)$, and (14) becomes

$$\Pr(E_I \,|\, Y \equiv 1) = \Pr(Z > X) = \int_0^1 \Pr(Z > x) dF_X(x) . \quad (15)$$

Now, if $\phi_{ji}$ is uniformly distributed, we then have

$$\Pr(Z > x) = \Pr(G^*_{T_j}(\phi_{ji}) > x)$$
$$= \int_0^{2\pi} 1_{\{\phi_{ji}:G^*_{T_j}(\phi_{ji})>x\}}(\phi_{ji}) d\phi_{ji}/2\pi = |\{\phi \,|\, G^*_{T_j}(\phi_{ji}) > x\}|/2\pi, \quad (16)$$

where $1_A(x)$ is the indicator function of set $A$, which is equal to 1 if $x \in A$, and 0 otherwise. That is, $\Pr(Z > x)$ is the normalized beam width of $G_{T_j}(\phi_{ji})$ with respect to threshold $x$ (the normalized distance between $T_j$ and $R_i$).

The physical meaning of (15) is straightforward: 1) the closer the potential interfering neighbor, the greater is the normalized beam width and the higher is the probability of interference; 2) the probability of interference is an expected value of normalized beam width with respect to the normalized distance. Thus, we define this expected value as the *effective beam width* of the antenna of $T_j$.

**Definition: Effective beam width:** The effective beam width of the antenna of $T_j$ is defined as follow:

$$W_B(T_j) \triangleq \Pr(Z > X) = \Pr(G^*_{T_j}(\phi_{ji}) > X) .$$

The corresponding *effective null width* is defined as its complement $\Pr(Z \leq X)$. They depend on the path loss exponent $\alpha$, the antenna pattern, the distribution of antenna orientation, and the distribution of the normalized distance

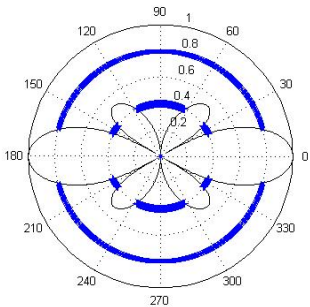

Figure 3. Null width of a particular pattern with respect to threshold 0.8 and 0.4 respectively. Null width is an increasing function of the threshold.

between our reference receiver and its interfering neighbor.

The effective beam width can also be calculated in another form. Note that $F_Z(1) = \Pr(Z \leq 1) = 1$ and $F_X(0) = 0$, (15) can be rewritten as

$$\begin{aligned}W_B(T_j) &= \Pr(E_I | Y \equiv 1) = \Pr(Z > X) = \int_0^1 \text{Prob}(Z > x)d(F_X(x)) \\ &= \int_0^1 [1 - F_Z(x)]d(F_X(x)) = [1-F_Z(x)]F_X(x)\Big|_0^1 + \int_0^1 F_X(x)d(F_Z(x)) \\ &= \int_0^1 F_X(x)d(F_Z(x)).\end{aligned} \quad (17)$$

*Directional Transmission and Directional Reception:*

Now let us investigate the scenario of directional transmission and directional reception. Define $\bar{x} = x/z$. Note that $F_X(x) = x^2 = \bar{x}^2 z^2 = F_X(\bar{x})F_X(z)$ and $0 \leq Y, Z \leq 1$. We have

$$\begin{aligned}\Pr(E_I) &= \Pr(YZ > X) \\ &= \int_0^1 \int_0^1 \Pr(Y > x/z | X = x, Z = z)d(F_X(x))d(F_Z(z)) \\ &= \int_0^1 \int_0^z \Pr(Y > x/z)d(F_X(x))d(F_Z(z)) \\ &= \int_0^1 \int_0^1 \Pr(Y > \bar{x})d(F_X(\bar{x}))F_X(z)d(F_Z(z)) \\ &= \Pr(Y > X)\int_0^1 F_X(z)d(F_Z(z)) \\ &= \Pr(Y > X)\Pr(Z > X) = W_B(R_i)W_B(T_j).\end{aligned} \quad (18)$$

Namely, the probability of interference is the product of effective beam widths of both the receiver and its interfering neighbor. In particular, assuming nodes are equipped with directional antennas of identical effective beam width $W_B$, the probability of interference in the case of directional transmission and directional reception is $W_B^2$. Compared with the case of directional transmission and omni-directional reception, this square law suggests a sharp reduction of interference.

*B. General Effective Beam Width and Its Properties*

The convenient property of interference probability in terms of effective beam width, the product form (18), is derived under the assumption of the "simplest case" where the interfering neighbor is uniformly distributed within the potential interference region with a uniform antenna orientation. Mathematically, we are assuming the following conditions:

C1) $\theta_{ij}$, $\phi_{ji}$ and $X$ are mutually independent;

C2) $F_X(x) = x^2$, $0 \leq x \leq 1$; $p(\theta_{ij}) = 1/2\pi$, $0 \leq \theta_{ij} < 2\pi$; $p(\phi_{ji}) = 1/2\pi$, $0 \leq \phi_{ji} < 2\pi$; .

However, the general definition of the *effective beam width* itself, $\Pr(Z > X)$ or $\Pr(Y > X)$, imposes no restriction on the distribution of $\theta_{ij}$, $\phi_{ji}$ and $X$. In fact, the distribution of $\theta_{ij}$, $\phi_{ji}$ and $X$ mainly depends on the active-node distribution (which is in turn determined by the underlying MAC protocol). The "simplest case" is just a specific scenario in which an ALOHA-like protocol and 2-Dimensional uniform distribution of nodes are assumed, as will be discussed in section IV. A question thus arises: is there a similar result to (18) that is applicable to a more general scenario? In the following discussion we will relax C2 while keeping C1.

We note that (17) is valid for general distributions of $X$ and $Z$ as long as $F_X(0) = 0$, which is a pretty mild restriction on $X$. Eq. (15) further requires that

$$F_X(\bar{x}z) = F_X(\bar{x})F_X(z), \forall\ 0 < \bar{x}, z \leq 1. \quad (19)$$

The non-constant continuous solutions of Cauchy equation $f(xy)=f(x)f(y)$ is $x^h$ ($h \neq 0$) [15]. Since our probability distribution function defined on [0, 1] should be non-decreasing, the continuous solution of functional equation (19) is a set of functions:

$$F_X(x) = x^h, (0 \leq x \leq 1, h \in \Re^+). \quad (20)$$

Hence, we can relax C2 to (20) while maintaining the validity of (18). Interestingly, Comparing (20) with C2, one can observe that 1) the product-form property requires no particular assumption on the distribution of the antenna orientation $\theta_{ij}$ and $\phi_{ji}$, so long as C1 is valid; 2) the distributions of $X$, $\theta_{ij}$ and $\phi_{ji}$ are still independent of $(1+\Delta)d_i$, so is the corresponding effective beam widths; and 3) for the normalized distance $X$, there are infinite many suitable distribution functions in (20) that preserve the product-form property, which can be view as a set of *basis distribution functions*, specified by the order $h$. (Particularly, our "simplest case" corresponds to the basis distribution function with $h = 2$.) We can also define $W_B^{(h)}(T_j) \triangleq \Pr(Z > X^{(h)})$ as the corresponding *basis effective beam width* of the antenna of $T_j$, where $F_{X^{(h)}}(x) = x^h, 0 \leq x \leq 1, h \in \Re^+$. In general, larger $h$ implies a heavier tail distribution of $X^{(h)}$, and that the interfering neighbor tends to be farther away from the reference receiver, and hence a smaller $W_B^{(h)}(T_j)$. This argument can be applied for a *rough* comparison of different node distribution and MAC protocols. For example, a random-access MAC protocol with *carrier sensing* may make sure that only nodes that are far apart can transmit together. So if a basis distribution function is used for the approximation in this case, it should have a larger $h$ than that in our "simplest case". Alternatively, we may also apply the similar technique as "Taylor expansion" to *accurately* model the scenario where a more general $F_X(x)$ is assumed.

Given C1, a general distribution of normalized distance $F_X(x)$ could be represented by a power series (e.g., Taylor series):

$$F_X(x) = \sum_h w_h x^h = \sum_h w_h F_{X^{(h)}}(x), 0 \leq x \leq 1, \quad (21)$$

where $\sum_h w_h = 1$ (according to the boundary condition $F_X(1) = 1$); or in a more general scenario, an integration of $x^h$:

$$F_X(x) = \int_0^{+\infty} w(h)x^h dh = \int_0^{+\infty} w(h)F_{X^{(h)}}(x)dh, 0 \leq x \leq 1, \quad (22)$$

where $\int_0^{+\infty} w(h)dh = 1$. These two cases represent a general spectrum of probability distribution $F_X(x)$ defined on [0, 1] (and hence a general spectrum of node distributions and MAC protocols). In fact, by simply letting $x = \exp(-s), s \in [0, +\infty)$, one can see that $F_X(e^{-s})$ in (22) is the one-side Laplace transform of the weight $w(h)$. Therefore, to get $w(h)$, we can simply take an inverse Laplace transform on $F_X(e^{-s})$. With (21) and (22), we could then have a more general version of (18). To see this, consider (21). By the definition of effective beam width, we have

$$W_B(T_j) = \int_0^1 \text{Prob}(Z > x) d(\sum_h w_h F_{X^{(h)}}(x)) \\ = \sum_h w_h \int_0^1 \text{Prob}(Z > x) d(F_{X^{(h)}}(x)) = \sum_h w_h W_B^{(h)}(T_j). \quad (23)$$

The probability of interference is

$$\Pr(E_I) = \Pr(YZ > X) \\ = \int_0^1 \int_0^z \Pr(Y > x/z) d(\sum_h w_h F_{X^{(h)}}(x)) d(F_Z(z)) \\ = \sum_h w_h \int_0^1 \int_0^1 \Pr(Y > \bar{x}) d(F_{X^{(h)}}(\bar{x})) F_{X^{(h)}}(z) d(F_Z(z)) \\ = \sum_h w_h \Pr(Y > X^{(h)}) \Pr(Z > X^{(h)}) = \sum_h w_h W_B^{(h)}(R_i) W_B^{(h)}(T_j). \quad (24)$$

Note that (24) is no longer a simple product of effective beam width as in (18), but a weighted sum of products of basis effective beam widths. The case of (22) is analogous. We just need to change the summation in (23) and (24) to an integration over $h$. We can prove that (see Appendix A), in both cases of (21) and (22), if $\forall h, w_h \geq 0, w(h) \geq 0$, then

$$\Pr(E_I) \geq W_B(R_i) W_B(T_j). \quad (25)$$

That is, the equality relationship in (18) becomes an inequality relationship in (25). Moreover, noting that $W_B^{(h)}(Node) \leq 1$, where $Node = T_j$ or $R_i$, by (23) and (24) we have

$$\Pr(E_I) \leq \min\{W_B(R_i), W_B(T_j)\}. \quad (26)$$

Combining the bounds in (25) and (26), we can bound the probability of interference with symmetric functions in terms of effective beam width at both sides, although we no longer have the equality relationship in (18). The conclusion that smaller effective beam width results in lower probability of interference is still valid even for a more general distribution of independent random variables $X$, $Y$ and $Z$.

In addition to the active-node distribution and antenna orientation, the path loss exponent $\alpha$ also affects the effective beam width. Specifically, $W_B(T_j)$ is a generally increasing function of $\alpha$. To see this, recall that $W_B(T_j) = \Pr(G_{T_j}(\phi_{ji}) > X^\alpha)$ and $0 \leq X \leq 1$. Given fixed antenna pattern $G_{T_j}(\phi_{ji})$ and joint distribution $F_{\phi_{ji}, X}(\phi, x)$, $\{G_{T_j}(\phi_{ji}) > X^{\alpha_1}\} \supseteq \{G_{T_j}(\phi_{ji}) > X^{\alpha_2}\}$ for $\alpha_1 > \alpha_2$. Hence $\Pr(G_{T_j}(\phi_{ji}) > X^{\alpha_1}) \geq \Pr(G_{T_j}(\phi_{ji}) > X^{\alpha_2})$. Therefore, a smaller path loss exponent $\alpha$ is favorable for interference cancellation. In particular, given the distribution of $G_{T_j}(\phi_{ji})$, the basis effective beam width with order $h$,

$$W_B^{(h)}(T_j) = \int_0^1 \Pr(G_{T_j}(\phi_{ji}) > x^\alpha) d(x^h) \\ = \int_0^1 \Pr(G_{T_j}(\phi_{ji}) > x^{\alpha/h}) d(x), \quad (27)$$

is an increasing function of *effective path loss exponent* $\alpha^* \triangleq \alpha/h$. Hence, for the impact on the *basis* effective beam width, adjusting the path loss exponent $\alpha$ is equivalent to inversely proportionally adjusting the order $h$.

Last but not least, one can easily extend the definition and properties of effective beam width to the scenario where *3-Dimensional* wireless networks are assumed. Particularly, let us consider the satellite network in space. In this particular scenario, we can adopt the free-space propagation model in which (2) is valid with a specific path loss exponent $\alpha = 2$. By choosing the spherical coordinate system [16], the normalized antenna gain $G(\theta_1, \theta_2)$ is now a function of azimuthal angle (longitude) $\theta_1$ and polar angle (colatitude) $\theta_2$. Hence, one can still use $\Pr(G^* > X)$ to define effective beam width. Its physical interpretation, for the case where the interfering neighbors are uniformly distributed in the potential interference region, is the total normalized *solid angle* in which the antenna gain $G^*$ is above the normalized distance $X$. In this case, by noting that a node with spherical coordinate $(x, \theta_1, \theta_2)$ is uniformly distributed within spherical region $x \leq 1$ iff $x$, $\theta_1$ and $\theta_2$ are mutually independent, and their probability density functions $p(x)$, $p(\theta_1)$ and $p(\theta_2)$ are

$$p(x) = 3x^2, 0 \leq x \leq 1; p(\theta_1) = 1/2\pi, 0 \leq \theta_1 < 2\pi; \\ p(\theta_2) = \sin \theta_2 / 2, 0 \leq \theta_2 \leq \pi;$$

one can observe that the distribution of radial part $x$ and angular part $\{\theta_1, \theta_2\}$ are mutually independent (which is similar to the condition C1), and (20) is valid with order $h = 3$ (or $\alpha^* = 2/3$). By simply repeating the derivation in the previous sections, one can prove that the product form in (18) is still valid even in 3-Dimensional space. Similarly, most of the properties established in this paper can be easily generalized to the 3-Dimensional scenario. In the following discussion, we will assume *planar* wireless networks unless otherwise specified.

*C. Numerical Scaling Law of Effective Beam Width of Some Particular Antenna Patterns*

The properties of the effective beam width with general antenna patterns and general distributions of $\theta_{ij}$, $\phi_{ji}$ and $X$ have been investigated in detail. This subsection examines the results of specific phased array antenna patterns in (8) and the scaling law of effective beam width in terms of $N$, the degree of array factor ($N+1$ is the number of elements in the antenna array). The motivation of this study is that the scaling law of the effective beam width is directly related to the scaling law of the network capacity, as will be shown. Unless otherwise specified, we adopt conditions C1 and C2 henceforth (see Subsection III.B). From the discussion in the previous section, again, this scaling law of effective beam width is independent of $d_i$ and

$\Delta$.

Consider the linear array. With respect to this family of array antenna patterns with the array factor as in (8), we first consider the case where the $N$ nulls of the array factor are positioned at $\theta_s = 2\pi s/(N+1)$ ($s = 1, 2, \ldots, N$, $N$ is even) and $D = \lambda/2$ in (9). By noting that $\sin\theta = \sin(\pi - \theta)$ in (9), one can easily identify the full set of nulls: $\{s\pi/(N+1), |s| = 1, 2, \ldots, N\}$. Moreover, due to the symmetry of the nulls' distribution, the maxima occur at angles $\{0, \pi\}$. Hence, the nulls are equally spaced in azimuthal angle between the two maxima. (The antenna pattern where $N = 4$ is shown in Fig. 2.) The corresponding effective beam width $W_B$ is now a function of the degree of array factor $N$ and the effective path loss exponent $\alpha^* = \alpha/h$. We can thus calculate the effective beam width $W_B$ in terms of $N$ for different $\alpha^*$ by Monte Carlo Integration [17] of (15).

The results are presented in Fig. 4. Notice that in most cases, the path loss exponent varies in the interval [1, 8] ([8]). Thus, the corresponding $\alpha^*$ belongs to interval [0.5, 4]. (Note that we fix $h$ to be 2. For a specific $N$, a general case of $\alpha^*$ shares the same $W_B$ with our case when $\alpha = 2\alpha^*$.) By varying $\alpha^*$ from 0.5 to 4, we find that $\lg W_B$ versus $\lg N$ is a bundle of straight lines which are almost parallel (see Fig. 4). Therefore there is a linear dependency between $\lg W_B$ and $\lg N$: $\lg W_B = -\gamma \lg N + b$, where the slope $\gamma$ (named as "beam-width decay index") is insensitive to $\alpha^*$ in our interested interval. On the other hand, the intercept $b$ is more sensitive to $\alpha^*$. It increases as $\alpha^*$ increases. This observation coincides with our previous analysis which shows that $W_B$ is an increasing function of $\alpha^*$. We can write

$$W_B = 10^b / N^\gamma = b_1 / N^\gamma. \quad (28)$$

That is, $W_B$ is inversely proportional to $N^\gamma$. This observation also agrees with work by others (e.g., [7]) which suggest that "more elements in the phased array antenna results in more nulls, sharper main beam and smaller side lobe on its pattern, and thus narrower beam widths", although the definitions of the various beam widths used by others (e.g., the half power beam width (HPBW), the first null beam width (FNBW), etc.) are different from that of our effective beam width. The name, beam-width decay index, for $\gamma$, also becomes clear from (28). It relates to how fast $W_B$ decays as $N$ increases. Larger $\gamma$ leads to better interference cancellation performance, and higher network capacity.

Now, for a typical value of $\alpha = 4$, by linear regression, we obtain $b_1$=0.659 and $\gamma$=0.810. For other $\alpha \in [1,8]$, $\gamma$ changes little. The insensitivity of $\gamma$ with respect to $\alpha^*$ also suggests that for a more general case of (21) and (22), $W_B$ (which is a weighted combination of basis effective beam widths of different $\alpha^*$) also roughly scales as $N^{-\gamma}$.

Similarly, when $\alpha^*$ is fixed (e.g., $\alpha^* = 2$), effective beam width $W_B$ is a function of $N$ and $D/\lambda$, where $D$ is the spacing between adjacent antenna elements. By varying $D$ from $\lambda/2$ to $\lambda/16$, we find that $\lg W_B$ versus $\lg N$ is also a bundle of straight lines which are almost parallel (see Fig. 5). Therefore, for different $D/\lambda$, the numerical scaling law (28) is still valid, with beam-width decay index $\gamma$ insensitive to $D/\lambda$ and intercept $b$ decreasing with $D/\lambda$.

More importantly, the numerical scaling law of (28) is not an exclusive property of the particular family of equally-spaced-null-linear-array (ESNLA) antenna patterns. In fact, additional simulation by us shows that the effective beam widths of many other linear phased array antenna patterns (e.g., the popular families of Binomial array and Chebyshev array [18]) share a similar form as (28). Specifically, the $(b_1, \gamma)$ pairs of Binomial array and Chebyshev array are (0.496, 0.496) and (0.716, 0.874) respectively, where $\alpha^* = 2$ and $D = \lambda/2$. (Note that the antenna pattern of Chebyshev array is actually specified by two parameters: $N$ and the main-lobe-to-side-lobe ratio $R_{MS}$. In our simulation we carefully select optimal $R_{MS}$ such that the effective beam widths of different $N$ are minimized.) We can rank these $W_B$ in a descending order (see Fig. 6):

$$Binomial \gg ESNLA \gtrsim Chebyshev,$$

which is also valid for other $\alpha^*$ and $D/\lambda$.

We believe that the numerical scaling law in the form of (28) is representative of other linear phased array antenna patterns. In the subsequent section, we will choose ESNLA as a representative of linear array. Although it has a slightly larger $W_B$ (or less $\gamma$) than the Chebyshev array, ENSLA has the advantage that careful optimization of $R_{MS}$ as in Chebyshev array is not necessary.

To summarize, we have defined the effective beam width as the interference probability of one-side directional transmission (reception). It lumps the impact of the channel path loss, antenna pattern and active-node distributions on network capacity into a single quantitative measure. The effective beam width contains several convenient properties for analytical purposes, (see (18), (24)-(28)). The definition of the effective beam width affords us a quantitative understanding (as opposed to qualitative, intuitive argument) of why and how a "good" antenna pattern usually consists of a sharp main beam, a good deal of nulls and small side lobes. We will show how effective beam width affects network capacity in the next section.

## IV. ASSESSMENT OF SCALING LAW OF NETWORK CAPACITY OF WIRELESS RANDOM NETWORKS WITH DIRECTIONAL ANTENNAS

We now analyze the impact of the effective beam width on the scaling law of network capacity [10].

### A. Random Network Model and Network Capacity

A wireless random network consists of a group of nodes $\mathcal{N} = \{1, 2, \ldots, n\}$, all of which are located randomly in certain area. Each node can act as a transmitter to form a directed link with its intended receiver within its transmission range and initiate a transmission over a wireless channel of constant bit rate $R$. We denote the full set of links by $\mathscr{S}$. For analytical convenience, we adopt the following assumptions:

A1) All nodes have identical transmission range $r$ which guarantees the (asymptotic) global connectivity [10];
A2) All transmissions are unicast and half-duplex;
A3) For a particular receiver, at most one packet can be successfully received at a single reception;
A4) Transmissions are slotted into synchronized time slots.

There are several variants for the definitions of network capacity that try to characterize the transport capability of a network in different ways. We focus on two of them here: maximal total throughput $C_{tt}$ and transport capacity $C_{tr}$. The *maximal total throughput $C_{tt}$* is defined as the maximum of total *bits per second* $\eta_{tt}$ that can be transported in the network. The *transport capacity $C_{tr}$* is the maximum of total transport throughput, *bit-distance product* per second $\eta_{tr}$, that can be transported in the network [10]. We will investigate the scaling

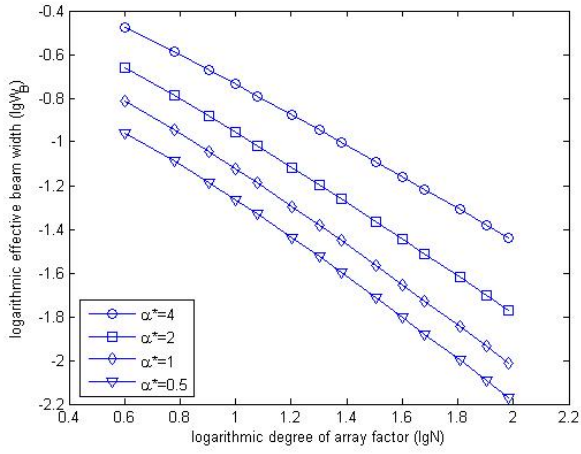

Figure 4. Logarithmic effective beam width (lg $W_B$) versus logarithmic degree of array factor (lg $N$) for different effective path loss exponenets $\alpha^*$, where $D = \lambda/2$ and equally-spaced-null linear array is assumed

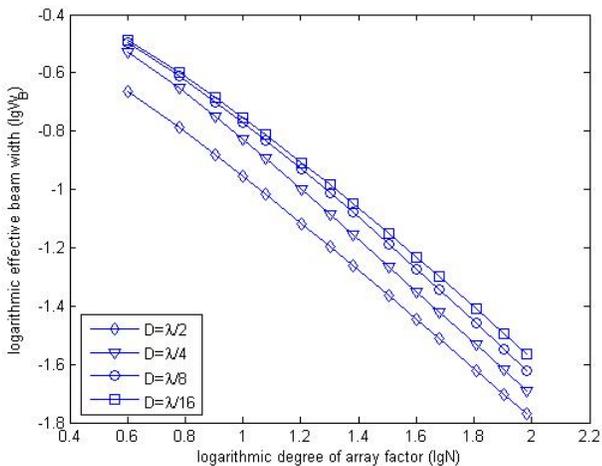

Figure 5. Logarithmic effective beam width (lg $W_B$) versus logarithmic degree of array factor (lg $N$) for different antenna-element spacing $D$, where $\alpha^* = 2$ and equally-spaced-null linear array is assumed

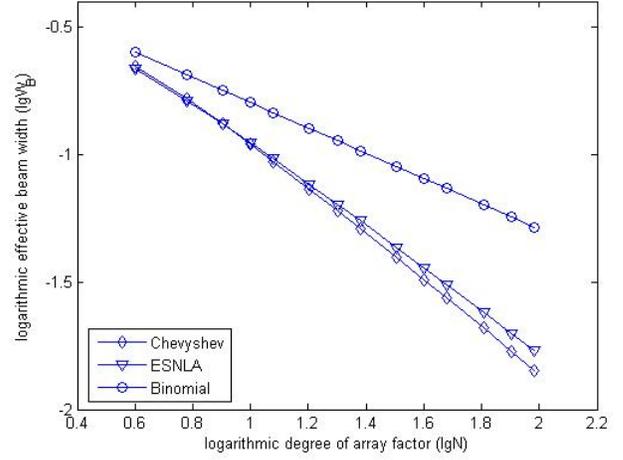

Figure 6. Logarithmic effective beam width (lg $W_B$) versus logarithmic degree of array factor (lg $N$) for different families of linear array, where $\alpha^* = 2$ and $D = \lambda/2$ is assumed

law with respect to these two measures as the network size, $n$, increases asymptotically. Thanks to the assumption of half duplexity, there can be at most $\lfloor n/2 \rfloor$ successful simultaneous transmissions at bit rate $R$ at any instant. Also, by noting that our unit torus is a wrap-around structure, the link length is bounded by half of the dimension of its diagonal $\sqrt{2}/2$. Therefore, we have

$$C_{tt} = O(n), C_{tr} = O(n) . \ddagger \qquad (29)$$

### B. Node Distribution and MAC Protocol

For simplicity, we assume all nodes are *independently* placed on the surface of a unit torus in a *uniformly* distributed manner. Each of them is equipped with an identical directional antenna. Moreover, we assume a slotted ALOHA-like random access scheme for transmission scheduling: in each time slot, each node will independently activate itself to be a transmitter with identical probability $p_t$. Once activated, a transmitter is *equally likely* to choose any neighbor within its transmission range to be its intended receiver. This is similar to the "simplest case" discussed in Subsection III.A. In the following discussion, we will use $W_B$ to denote the particular effective beam width under that "simplest case" (where C1 and C2 is valid).

We start with the simple scenario of *directional transmission and omni reception* in Subsections C and D. In this case, a transmitter $T_i$ will steer its main beam to its intended receiver $R_i$

---

‡ We adopt the following notations to represent asymptotic bounds:

1. $f(n) = O(g(n))$ implies there exists certain constant $c$ and integer $N$ such that $f(n) \leq cg(n)$ for $n > N$.

2. $f(n) = \Omega(g(n))$ implies $g(n) = O(f(n))$.

3. $f(n) = \Theta(g(n))$ implies $f(n) = O(g(n))$ and $g(n) = O(f(n))$.

The bounds for random networks hold with high probability (i.e., they hold with probability 1 when $n \to \infty$).

to form link $i$. Each receiver will use an omni-directional antenna to listen and receive data. *Directional transmission and directional reception* will be treated in Subsection E. Subsection F will generalize the result to the scenario where multiple-interference model and Rayleigh fading channel are assumed.

### C. Probability of Transmission to be Success and Per-Link (Transport) Throughput

In the scenario of *directional transmission and omni reception*, the transmission of link $i$ is considered to be a success if and only if 1) its receiver $R_i$ is not transmitting and 2) the reception at $R_i$ is interference-free with respect to the remaining $n$-2 nodes. Condition 2) can be decomposed into the intersection of a series of events $\bigcap_{j \in \mathcal{N}, j \neq T_i, R_i} E_{INT}^C(j,i)$, where $E_{INT}(j,i)$ is the event that {*node $j$ interferes with the transmission of link $i$*}, and the superscript $C$ denotes the complement of event. We have $\Pr(E_{INT}(j,i)) = \Pr(E_{INT}(j,i) \cap E_T(j,i)) + \Pr(E_{INT}(j,i) \cap E_T^C(j,i))$, where $E_T(j,i)$ is the event {*node $j$ intends to transmit to $R_i$*}. Let $k_{pr}(j)$ be the number of potential receivers of a node $j$. Consider the case where $(1+\Delta)d_i < r$. By chain rule of conditional probability, we can write $\Pr(E_{INT}(j,i) \cap E_T^C(j,i)) = \Pr(E_1)\Pr(E_2)\Pr(E_3)$, where events $E_1$ = {*node $j$ is an active transmitter*}, $E_2$ = {*node $j$ is a potential interfering neighbor of $R_i$, but not transmitting to $R_i$*} and $E_3$ = {*node $j$'s transmission to another node other than $R_i$, interferes with reception at $R_i$*}. By the previous assumption, whether node $j$ is a transmitter or not is independent of its location. Moreover, node $j$ is uniformly distributed within a unit torus. We have $\Pr(E_1) = p_t$ and $\Pr(E_2) = c_1 d_i^2[1 - 1/k_{pr}(j)]$, where $c_1$ is defined as $\pi(1+\Delta)^2$. (For a given $SIR_0$ and $\alpha$, the guard zone $\Delta = SIR_0^{1/\alpha} - 1$ is a constant, so is $c_1$. For example, assuming a typical value of $SIR_0 = 10$ and $\alpha = 4$, we have $\Delta = 0.778$ and $c_1 = 9.935$. Both of them will not affect the order of network capacity, as will be shown shortly.) Event $E_3$ is the same as event $E_1$ of the "simplest case" in section III.A. Hence, for directional transmission and omni-directional reception, $\Pr(E_3) = W_B$. For our case of $(1+\Delta)d_i < r$, event $E_{INT}(j,i) \cap E_T(j,i)$ is exactly {*transmitter node $j$ is a potential interfering neighbor transmitting to $R_i$*}. Therefore, we have $\Pr(E_{INT}(j,i) \cap E_T(j,i)) = p_t c_1 d_i^2 / k_{pr}(j)$. Then $\Pr(E_{INT}^C(j,i)) = 1 - p_t c_1 d_i^2 [W_B + (1-W_B)/k_{pr}(j)]$. One can prove

$$[1 - p_t \pi r^2 / k_{pr}(j)](1 - c_1 p_t d_i^2 W_B)$$
$$\leq \Pr(E_{INT}^C(j,i)) \leq 1 - c_1 p_t d_i^2 W_B, \quad (30)$$

which is also valid for the case of $(1+\Delta)d_i \geq r$. To conserve space, we omit the proof here. Due to the independent location and transmission behavior of each node, from our pair-wise interference model (6), events $E_{INT}(j,i)$ ($\forall j \in \mathcal{N}, j \neq T_i, R_i$) are mutually independent. Thus, the probability of the transmission of link $i$ to be success, $p_{suc}(i)$ can be written as a product form

$$p_{suc}(i) = (1 - p_t) \prod_{j \in \mathcal{N}, j \neq T_i, R_i} \Pr(E_{INT}^C(j,i)). \quad (31)$$

The first term of the product is the probability of $R_i$ not transmitting. We can prove that (see Appendix B)

$$p_{suc}(i) = \Theta\{(1 - p_t)(1 - p_t c_1 d_i^2 W_B)^{n-2}\}. \quad (32)$$

From (32), we can calculate the per-link throughput. According to the transmission protocol, in each time slot, node $j$ will transmit to a particular potential receiver with probability $p_t / k_{pr}(j)$. Hence, in a long run, the link throughput $\eta(i)$ and link transport throughput $\eta_{tr}(i)$ of link $i$ are given by

$$\eta(i) = R p_t p_{suc}(i) / k_{pr}(T_i)$$
$$= \Theta\{p_t(1 - p_t)(1 - p_t c_1 d_i^2 W_B)^{n-2} / k_{pr}(T_i)\}; \quad (33)$$

$$\eta_{tr}(i) = d_i \eta(i) = \Theta\{d_i p_t(1 - p_t)(1 - p_t c_1 d_i^2 W_B)^{n-2} / k_{pr}(T_i)\}. \quad (34)$$

### D. Scaling Law of Network Capacity

The total throughput is the summation of per-link throughput over all link $i$ ($i \in \mathcal{S}$). We consider the *expected total throughput* $\eta_{tt}$ and the *expected total transport throughput* $\eta_{tr}$ by averaging over the node distribution. Recall that for a link $i \in \mathcal{S}$, due to the uniform node distribution and random selection of $R_i$ within the transmission range of $T_i$, the probability density function of link length $d_i$ also follows (13), where $\rho = d_i$ and $R_0 = r$. Thus, we can figure out $\eta_{tt}$ and $\eta_{tr}$ as follows:

$$\eta_{tt} = \sum_{T_i \in \mathcal{S}} k_{pr}(T_i) E_{d_i}[\eta(i)] = n E_{d_i}[k_{pr}(T_i)\eta(i)]$$
$$= \Theta\{\int_0^r [2np_t(1-p_t)(1-c_1p_td_i^2W_B)^{n-2}d_i/r^2]d(d_i)\} \quad (35)$$
$$= \Theta\{n(1-p_t)[1-(1-c_1p_tr^2W_B)^{n-1}]/[c_1(n-1)r^2W_B]\};$$

$$\eta_{tr} = \sum_{T_i \in \mathcal{S}} k_{pr}(T_i) E_{d_i}[\eta_{tr}(i)] = n E_{d_i}[k_{pr}(T_i)d_i\eta(i)]$$
$$= \Theta\{\int_0^r [2np_t(1-p_t)(1-c_1p_td_i^2W_B)^{n-2}d_i^2/r^2]d(d_i)\}. \quad (36)$$

Both $\eta_{tt}$ and $\eta_{tr}$ are functions of $p_t$, $r$ and $W_B$. We note that the quantities in (32)-(36) are decreasing functions of $W_B$. Namely, smaller $W_B$ yields better performance. As for System parameters $p_t$ and $r$, there may be tradeoffs. For instance, higher $p_t$ leads to more active transmissions attempts, but will also creates more interference. Larger $r$ increases the average link length in the transport throughput, but also results in more interfering neighbors; and this may in turn pull down $\{p_{suc}(i)\}_{i \in \mathcal{S}}$. For a given $W_B$, the "order" of the corresponding network capacity by optimal selection of $p_t$ and $r$ in (35) and (36) are summarized below (see Appendix C for constraint on the region of optimality for $p_t$ and $r$):

The scaling law of maximal total throughput $C_{tt}$ is given by:

$$C_{tt} = \Theta(nW_B^{-1}\log^{-1}(n)), \text{ if } W_B = \Omega(\log^{-1}(n));$$
$$C_{tt} = \Theta(n), \text{ if } W_B = O(\log^{-1}(n)). \quad (37)$$

with the selection of system parameters $r = \Theta(\sqrt{\log(n)/n})$ and $p_t = 1/2$ in (35) (see Appendix D).

The scaling law of transport capacity $C_{tr}$ can also be calculated as (see Appendix E)

$$C_{tr} = \Theta(W_B^{-1} n^{1/2} \log^{-1/2}(n)), \text{ if } W_B = \Omega(\log^{-1}(n));$$
$$C_{tr} = \Theta(W_B^{-1/2} n^{1/2}), \text{ if } W_B = O(\log^{-1}(n)) \text{ and } \Omega(n^{-1}); \quad (38)$$
$$C_{tr} = \Theta(n), \text{ if } W_B = O(n^{-1}),$$

where $p_t$ and $r$ are chosen as

$$p_t = \Theta(\log^{-1}(n) W_B^{-1}), r = \Theta(\sqrt{\log(n)/n}), \text{ if } W_B = \Omega(\log^{-1}(n));$$
$$p_t = 1/2, r = \Theta(W_B^{-1/2} n^{-1/2}), \text{ if } W_B = O(\log^{-1}(n)) \text{ and } \Omega(n^{-1}). \quad (39)$$

### E. Directional Transmission and Directional Reception

The analysis for the case of directional transmission and directional reception is similar except that 1) the interference probability becomes $\Pr(E_3) = W_B^2$; and 2) in addition to mutual interference, we need to consider "*conflict*" in transmission as well. Node $j$ is said to be in *conflict* with link $i$ if $j$ and $T_i$ transmit packets to $R_i$ simultaneously. In the case of omni-directional reception, the above conflict is simply handled by treating node $j$ as an interfering neighbor with $\phi_{ji} = 0$ in (4), because the receiver never needs to steer its main beam. However, in the case of directional reception, a receiver with more than one transmitter in a time slot must decide how to steer its beam.

In practice, we assume that there is an underlying antenna steering protocol to control the main beam orientation of our receiver antenna. For example, a receiver antenna may sweep its main beam in a circular manner to detect and receive training sequences from intended transmitters at the beginning of a time slot. As an alternate scheme, the receiver may also use an omni-directional antenna at the beginning of a time slot to go through a handshake procedure, during which the transmitter trains the receiver to steer its beam toward the transmitter. The handshake control data could be transmitted at a lower rate than the regular data to deal with the lower SIR associated with the use of omni reception. Once the direction of the receiver is trained, the regular data can then be transmitted at higher speed. There can be other schemes also. The focus of this paper is not on these beam training protocols and we will not dwell on further details here. With the assumption of one of the schemes, if multiple transmitters send control data to the same receiver, and none succeeds in capturing the receiver, then there will be a collision. Although there is still the possibility that one of the transmitters will succeed in capturing the receiver in case of a conflict (say, because it is physically closer to the receiver), as a conservative analysis, we assume no such capturing in the following. It can be shown that this conservative assumption does not affect the order of our result.

Again we consider the probability of a transmission being successful. The transmission of link $i$ is successful if and only if 1) its receiver $R_i$ is not transmitting and 2) the reception at $R_i$ is interference-free and *conflict-free* with respect to the remaining $n$-2 nodes. Condition 2) can be decomposed to the intersection of a series of independent events $\bigcap_{j \in \mathcal{N}, j \neq T_i, R_i} E_{IF\&CF}(j,i)$, where $E_{IF\&CF}(j,i)$ is the event {*link $i$ is interference-free (IF) and conflict-free (CF) from node $j$*}. We again consider the case where $(1+\Delta)d_i < r$. From set theory, we have $\Pr(E_{IF\&CF}(j,i)) = 1 - \Pr(E_{CON}(j,i)) - \Pr(E_{INT\&CF}(j,i))$, where $E_{CON}(j,i)$ is the event {*node $j$ conflicts with the transmission of link $i$*}, which is the intersection of the following two independent events {$j$ is located within the transmission range of $R_i$} and {$j$ is a transmitter intended for $R_i$}. Hence, we have $\Pr(E_{CON}(j,i)) = p_t \pi r^2 / k_{pr}(j)$. We can also prove that $\Pr(E_{INT\&CF}(j,i)) = p_t \pi (1+\Delta)^2 d_i^2 W_B^2 [1 - 1/k_{pr}(j)]$. Similar to (30), we have

$$[1 - p_t \pi r^2 / k_{pr}(j)](1 - c_1 p_t d_i^2 W_B^2) \\ \leq \Pr(E_{IF\&CF}(j,i)) \leq 1 - c_1 p_t d_i^2 W_B^2, \quad (40)$$

which holds even for the case of $(1+\Delta)d_i \geq r$. Similarly follows the step from (30) to (32), we have

$$p_{suc}(i) = \Theta\{(1-p_t)(1 - p_t c_1 d_i^2 W_B^2)^{n-2}\}, \quad (41)$$

which can be viewed as simply replacing the $W_B$ in (32) by $W_B^2$. From a similar derivation, we can get similar results for network capacity, by replacing $W_B$ in (37)-(39) with $W_B^2$ also. This square law suggests a significant improvement in network capacity. For example, the maximal $W_B$ required for $\Theta(n)$ scalability of $C_{tr}$ improves from $\Theta(n^{-1})$ to $\Theta(n^{-1/2})$.

By assuming phased array antennas pattern as in (8), we can also substitute the numerical scaling law (28) into the scaling law of network capacity. Specifically, for the scenario where $W_B = \Omega(\log^{-1/2}(n))$ and $\gamma = 0.810$ (ESNLA), one can observe a substantial improvement (at a factor of $\Theta(N^{2\gamma}) = \Theta(N^{1.620})$) in $C_{tr}$ over that with omni-directional antennas (where $W_B \equiv 1$).

### F. Generalization under Multiple-Interference Model and Rayleigh Fading Channel

So far, based on the pair-wise-interference model (5) and no-fading assumption, the properties of effective beam width and its impact on network scalability have been investigated in detail. However, as has been discussed in Subsection II.B, pair-wise-interference model (5) is just a simplified and relaxed variant of our multiple-interference model (6), leading to an upper bound estimation of network capacity. One may question whether effective beam width and the corresponding scaling law (37), (38) is still applicable in the more realistic scenario, where multiple-interference model and channel fading is taken into account. We address this issue in the context of *Rayleigh fading*, showing that the answer is "yes".

We will first identify the inherent relationship between the following two scenarios:

S1) pair-wise-interference model with no fading; and
S2) multiple-interference model with Rayleigh fading;
paving the way for our generalization.

In the scenario S2, we also adopt the setting and assumption in Subsections A and B. Since the model has been changed from (6) to (5), the major difference between these two scenarios lies in the condition of successful transmission. Let $E_{sur}(i)$ denote the event {link $i$ can survive the cumulative interference from all other nodes}. Then for scenario S2, the transmission of link $i$ is considered to be a success if and only if 1) its receiver $R_i$ is not transmitting and 2) $E_{sur}(i)$ occurs. We are interested in the probability $\Pr(E_{sur}(i))$. Similar to the discussion in Subsection II.B, the signal power received at $R_i$ is given by $S_i = K \cdot F_{R_iT_i} \cdot P / d_i^\alpha$, while the interference power from *node k* to link $i$ (denoted by $I_{ik}$) is given by $I_{ik} = K \cdot F_{R_ik} \cdot P \cdot \zeta_k \cdot G_{R_i}(\theta_{ik}) \cdot G_k(\phi_{ki}) / |k - R_i|^\alpha, k \in \mathcal{N}, k \neq T_i, R_i$, where $\zeta_k$ are i.i.d. Bernoulli random variables with parameter $p_t$, which models the independent transmission behavior of each node $k$; antenna orientation $\theta_{ik} = \angle T_i R_i k, \phi_{ki} = \angle R(k) k R_i$, where $R(k)$ is the intended receiver of node $k$ once it is activated.

One can observe the slight difference between the $I_{ik}$ above and the $I_{ij}$ in Subsection II.B, i.e., $k$ is a node while $j$ is an active link. The motivation for our "redefinition" of $I_{ik}$ here is to facilitate the subsequent analysis.

According to the multiple-interference model, $E_{sur}(i)$ occurs iff

$$SIR_i = S_i / \sum_{k \in \mathcal{N}, k \neq T_i, R_i} I_{ik} \geq SIR_0, \quad (42)$$

Recall that in Rayleigh fading scenario, $F_{R_ik}(k \in \mathcal{N}, k \neq R_i)$ are independent exponentially distributed random variables with unit mean. We can use $\overline{S_i}$ ($\overline{I_{ik}}$) to denote the expected value of $S_i$ ($I_{ik}$) taken over the fading state $F_{R_iT_i}$ ($F_{R_ik}$):

$$\overline{S_i} = K \cdot P / d_i^\alpha;$$
$$\overline{I_{ik}} = K \cdot P \cdot \zeta_k \cdot G_{R_i}(\theta_{ik}) \cdot G_k(\phi_{ki}) / |k - R_i|^\alpha.$$

Then we have $S_i = F_{R_iT_i} \overline{S_i}$ and $I_{ik} = F_{R_ik} \overline{I_{ik}}$. By noting that $F_{R_iT_i}$, $F_{R_ik}$ and $\overline{I_{ik}}$ are mutually independent, we have

$$\Pr(E_{sur}(i)) = \Pr(F_{R_iT_i} \overline{S_i} \geq SIR_0 \sum_{k \in \mathcal{N}, k \neq T_i, R_i} I_{ik})$$
$$= E_{\{\overline{I_{ik}}\}, \{F_{R_ik}\}}[\Pr(F_{R_iT_i} \overline{S_i} \geq SIR_0 \sum_{k \in \mathcal{N}, k \neq T_i, R_i} I_{ik} | \{\overline{I_{ik}}\}, \{F_{R_ik}\})]$$
$$= E_{\{\overline{I_{ik}}\}, \{F_{R_ik}\}}[\exp(-SIR_0 \sum_{k \in \mathcal{N}, k \neq T_i, R_i} I_{ik} / \overline{S_i}) | \{\overline{I_{ik}}\}, \{F_{R_ik}\}]$$
$$= \prod_{k \in \mathcal{N}, k \neq T_i, R_i} E_{\overline{I_{ik}}, F_{R_ik}}[\exp(-SIR_0 I_{ik} / \overline{S_i}) | \overline{I_{ik}}, F_{R_ik}]$$
$$= \prod_{k \in \mathcal{N}, k \neq T_i, R_i} \Pr(S_i / I_{ik} \geq SIR_0).$$

Hence the probability of success for the transmission of link $i$, $p_{suc}(i)$, can still be expressed in a product form like (31):

$$p_{suc}(i) = (1 - p_t) \prod_{k \in \mathcal{N}, k \neq T_i, R_i} \Pr(S_i / I_{ik} \geq SIR_0). \quad (43)$$

By carefully comparing (43) to (31), one can find that, the expressions of $p_{suc}(i)$ in two different scenarios S1 and S2 appear to be exactly the same! In fact, one can easily verify in the no-fading scenario (where $F_{R_ik} \equiv 1(k \in \mathcal{N}, k \neq R_i)$) that

$$\Pr(E_{INT}^C(k,i)) = \Pr(\overline{S_i} / \overline{I_{ik}} \geq SIR_0) = \Pr(S_i / I_{ik} \geq SIR_0). \quad (44)$$

The fundamental relationship between (43) and (31) is attributed to the nice properties of exponential distribution and exponential function (originated form Rayleigh fading). They transform the cumulative multiple interference into a product form which *appears to be* the same as its pair-wise no-fading counterpart. The sole difference is that $F_{R_ik}(k \in \mathcal{N}, k \neq R_i)$ are no longer unit constant, but independent exponentially distributed random variables with unit mean.

Therefore, according to (43) and (44), we only need to pay attention to the difference in $\Pr(S_i / I_{ik} \geq SIR_0)$ for the two scenarios S1 and S2. Again consider the simple case of directional transmission and omni reception. We have

$$\Pr(S_i / I_{ik} \geq SIR_0)$$
$$= E_{F_{R_iT_i}, F_{R_ik}} \{\Pr(\overline{S_i} / \overline{I_{ik}} \geq [SIR_0(F_{R_ik} / F_{R_iT_i})] | F_{R_iT_i}, F_{R_ik})\} \quad (45)$$
$$= E_{F_{R_iT_i}, F_{R_ik}} \{\Pr(E_{INT}^C(k,i)) |_{SIR_0 \text{ replaced by } [SIR_0(F_{R_ik}/F_{R_iT_i})]}\}.$$

From (30), we have

$$[1 - p_t \pi r^2 / k_{pr}(k)](1 - c_1 p_t d_i^2 W_B) \leq \Pr(E_{INT}^C(k,i))$$
$$= \Pr(\overline{S_i} / \overline{I_{ik}} \geq SIR_0) \leq 1 - c_1 p_t d_i^2 W_B. \quad (46)$$

By combining (45) and (46) we can get

$$[1 - p_t \pi r^2 / k_{pr}(j)](1 - c_1 F(\alpha) p_t d_i^2 W_B)$$
$$\leq \Pr(E_{INT}^C(j,i)) \leq 1 - c_1 F(\alpha) p_t d_i^2 W_B, \quad (47)$$

where $F(\alpha) = E[(F_{R_ik} / F_{R_iT_i})^{2/\alpha}]$ is a function of path loss exponent $\alpha$. It can be shown that (see Appendix F)

$$F(\alpha) = \begin{cases} +\infty, \alpha \leq 2; \\ [\sin c(2\pi/\alpha)]^{-1}, \alpha > 2. \end{cases} \quad (48)$$

where $sinc(x) = \sin(x)/x$. When $\alpha > 2$ is fixed, $F(\alpha)$ is a constant greater than one which can be absorbed into the constant $c_1$ (For instance, when a typical value $\alpha = 4$ is assumed, $F(\alpha) = \pi / 2$.). This will lead to a reduction in $p_{suc}(i)$, but its order as in (32) remains. Therefore, by simply repeating the former derivation, we can obtain the same order result in (37)-(39). The case of directional transmission and directional reception is similar.

To summarize, thanks to the convenient property of Rayleigh fading, the definition of effective beam width and its impact on the *order* of network capacity preserves, even when the more realistic scenario S2 is under consideration.

## V. CONCLUSION

The investigations in this paper have been a first attempt to *quantitatively* capture the characteristic of directional antennas that are responsible for their network-capacity boosting capability. Our main contributions are as follows:

1. We have introduced the concept of the *effective beam width*. We point out that the capacity-boosting capability of directional antennas is not due to their "isolated" characteristics alone. Rather, it is due to the combined effects of (i) antenna pattern, (ii) active-node distribution, and (iii) channel characteristic. These effects are lumped together into the single effective beam width measure.

2. We have investigated the mathematical properties of *effective beam width,* and demonstrate how to apply these convenient properties to analyze network performance. Interestingly, we find that the probability of interference is the product of effective beam widths of the receiver and its interfering neighbor, under a rather mild condition C1 and (20). We have also shown that a phased array antenna with $N$ elements can boost transport capacity of an Aloha-like network by a factor of $\Theta(N^{1.620})$.

3. We have presented a fundamental relationship which ties the multi-user interference model with Rayleigh fading to the pair-wise-interference model with no fading. This relationship preserves the definition and properties of effective beam width and the order of network capacity in both scenarios. This is an interesting intellectual result for the following reason. Although the pair-wise-interference model has been commonly adopted in the research community (primarily to ensure analytical tractability), it is usually viewed as a simplified and approximated version of multiple-interference model. Its validity has often been challenged. Our results broaden the applicable scenarios of the pair-wise interference model.

## APPENDIX

### A. Proof of equation (25)

*Proof*: For simplicity, we only consider the case of (21). The case of (22) is similar. Rewrite (23) and (24) as

$$W_B(T_j) = \sum_{i=1}^{H} w_{h_i} W_B^{(h_i)}(T_j), \text{ and } \Pr(E_1) = \sum_{i=1}^{H} w_{h_i} W_B^{(h_i)}(R_i) W_B^{(h_i)}(T_j),$$

where the series $\{w_{h_i}\}_{i=1,\ldots,H}$ is positive, and the real series $\{h_i\}_{i=1,\ldots,H}$ is monotonically increasing. Consequently, the series $\{W_B^{(h_i)}(Node)\}_{i=1,\ldots,H}$ is monotonically decreasing, where $Node = T_j$ or $R_i$. We then have

$$\Pr(E_1) - W_B(R_i) W_B(T_j)$$
$$= [\sum_{k=1}^{H} w_{h_k} W_B^{(h_k)}(R_i) W_B^{(h_k)}(T_j)] \sum_{k=1}^{H} w_{h_k}$$
$$- [\sum_{k=1}^{H} w_{h_k} W_B^{(h_k)}(R_i)][\sum_{k=1}^{H} w_{h_k} W_B^{(h_k)}(T_j)]$$
$$= \sum_{1 \le k, g \le H, k \ne g} w_{h_k} w_{h_g} \{[W_B^{(h_k)}(R_i) - W_B^{(h_g)}(R_i)]$$
$$[W_B^{(h_k)}(T_j) - W_B^{(h_g)}(T_j)]\} \ge 0.$$

The equality holds if and only if $H=1$, which corresponds to the case of *basis* effective beam width; or $\Pr(G^* \notin \{0,1\}) = 0$ (where $G^* = Y$ or $Z$), for which a typical case is that in which the antenna pattern $G^*(\theta)$ is consisting of sector beams. □

### B. Proof of equation (32)

*Proof*: From (30) to (32), we only need to prove that
$$M \triangleq \prod_{j \in \mathcal{I}, j \ne T_i, R_i} (1 - p_t \pi r^2 / k_{pr}(j)) = \Theta(1).$$

It is obvious that $M < 1$. Note that $k_{pr}(j) = \Theta(n\pi r^2)$ with high probability [19]. Thus, there exists $c_4$ such that $c_4 \ge n\pi r^2 / k_{pr}(j)$. Notice that $0 \le p_t \le 1$, we have
$$1 > M \ge (1 - p_t c_4 / n)^{n-2} = e^{-c_4 p_t} \ge e^{-c_4} = \Theta(1).$$
Therefore $M = \Theta(1)$. □

### C. Constraint on Region of optimality for pt and r(n)

To guarantee (asymptotic) connectivity of the whole network, $r$ should be $\Omega((\log(n)/n)^{1/2})$ [10]. In addition, the optimal transmitting probability $p_t \le 1/2$. For simplicity, we only consider (33) here. Let
$$\eta_1(i) \triangleq p_t(1 - p_t)(1 - p_t c_1 d_i^2 W_B)^{n-2}. \quad (49)$$
For any $p_{t0} \in (1/2, 1)$, by simply substituting $p_t = p_{t0}$ and $p_t = 1 - p_{t0}$ in (49), one can see that $\eta_1(i)|_{p_t=p_{t0}} < \eta_1(i)|_{p_t=1-p_{t0}}$. Hence the optimal $p_t$ should occur in $(0, 1/2]$. The proof for other objectives is similar.

### D. Scaling Law of Maximal Total Throughput

Note that we can restrict $p_t \le 1/2$ and hence $1/2 \le 1 - p_t \le 1$, $1 - p_t = \Theta(1)$. For asymptotic $n$, (35) becomes
$$\eta_{tt} = \Theta\{W_B^{-1} r^{-2} [1 - (1 - c_1 p_t r^2 W_B)^{n-1}]\}. \quad (50)$$
We define $f_1(r, p_t) = W_B^{-1} r^{-2} [1 - (1 - c_1 p_t r^2 W_B)^{n-1}]$. It is increasing on $p_t$. Hence, to maximize the order of $\eta_{tt}$, we have $p_t = 1/2$. Then $f_1(r, 1/2)$ is a decreasing function of $r$. Therefore the optimal $r = \Theta(\sqrt{\log(n)/n})$. Substituting the optimal $p_t$ and $r$ into (50), we get the maximal total throughput
$$C_{tt} = \Theta\{[1 - (1 - \Theta(\log(n) W_B / n))^{n-1}] W_B^{-1} n \log^{-1}(n)\}.$$
Recall that $\lim_{n \to +\infty}(1 - c_3/n)^n = \exp(-c_3)$. If the effective beam width $W_B = \Omega(\log^{-1}(n))$, then
$$\lim_{n \to +\infty}(1 - \Theta(\log(n) W_B / n))^{n-1} = \exp(-\Theta(\log(n) W_B)) = O(1).$$
Hence, $C_{tt} = \Theta(n W_B^{-1} \log^{-1}(n))$. Particularly, if $W_B = \Theta(\log^{-1}(n))$, we have $C_{tt} = \Theta(n)$, which implies that the order upper bound of maximal total throughput (29) is achieved. Although an even smaller $W_B$ yields a better leading coefficient before the order, the order of $C_{tt}$ can not be further improved.

### E. Scaling Law of Transport Capacity

Consider the integration in (36). Although its closed form cannot be obtained, we can study the lower bound $\eta_{trl}$ and upper bound $\eta_{tru}$ of total transport throughput. Since $d_i \le r$, by (30), (31) and (32), we have
$$\eta_{tr} \le nE_{d_i}[k_{pr}(T_i) r \eta(i)] = r \eta_{tt}$$
$$= \Theta\{W_B^{-1} r^{-1} [1 - (1 - c_1 p_t r^2 W_B)^{n-1}]\} \triangleq \eta_{tru}.$$

We define $f_2(r, p_t) = W_B^{-1} r^{-1}[1-(1-c_1 p_t r^2 W_B)^{n-1}]$. It is still an increasing function of $p_t$. Hence, the optimal $p_t$ is still 1/2. Then $f_2(r, 1/2)$ is a function of $r$. It *increases* on interval $(0, r_0)$ and then *decreases*. Here $r_0$ is the solution of $\partial f_2(r,1/2)/\partial r = 0$. Equivalently, we have $2(n-1)w(1-w)^{n-2} = 1-(1-w)^{n-1}$, where $w = c_1 W_B r_0^2 / 2$. The asymptotic solution of $w$ in $(0, 1)$ is $1.256/n$. Therefore, we have $r_0 = \sqrt{2*1.256/c_1 W_B n} = \Theta(W_B^{-1/2} n^{-1/2})$. If $W_B$ is $\Omega(\log^{-1}(n))$, since $r \geq \Theta(\sqrt{\log(n)/n}) \geq r_0$, the optimal $r$ is $\Theta(\sqrt{\log(n)/n})$. If $W_B$ is $O(\log^{-1}(n))$, the optimal $r$ is $r_0$. Hence, we can figure out the upper bound of transport capacity $C_{tru}$, which is exactly the same as (38).

On the other hand, we have a lower bound for (36):
$$\eta_{tr} \geq \int_0^r [2np_t(1-p_t)(1-c_1 p_t r^2 W_B)^{n-2} d_i^2 / r^2] d(d_i)$$
$$= 2np_t(1-p_t)r(1-c_1 p_t r^2 W_B)^{n-2}/3$$
$$= \Theta\{np_t r(1-c_1 p_t r^2 W_B)^{n-2}\} \triangleq \eta_{trl}.$$

Substituting $p_t$ and $r$ as (39) into $\eta_{trl}$, the scaling law of transport capacity in (38) follows, since the order of lower bound matches that of upper bound.

*F. Proof of equation (48)*

*Proof*: Define $V = F_{R_i k} / F_{R_i T_i}$, which is the ratio of two independent exponentially distributed random variables with unit mean. Its probability density function is given by $p_V(v) = (1+v)^{-2}$, $0 < v < +\infty$ (we omit the proof here). Then
$$F(\alpha) = E[(F_{R_i k} / F_{R_i T_i})^{2/\alpha}] = E[V^{2/\alpha}] = \int_0^{+\infty} v^{2/\alpha} (1+v)^{-2} dv.$$

For the case of $\alpha \leq 2$, we have
$$\int_0^{+\infty} v^{2/\alpha}(1+v)^{-2} dv > \int_1^{+\infty} v(1+v)^{-2} dv \to +\infty.$$

For the case of $\alpha > 2$, since $F_{R_i k}$ and $F_{R_i T_i}$ are independent, we have
$$F(\alpha) = E[(F_{R_i k} / F_{R_i T_i})^{2/\alpha}] = E[F_{R_i k}^{2/\alpha}] E[F_{R_i T_i}^{-2/\alpha}]$$
$$= \int_0^{+\infty} x^{2/\alpha} e^{-x} dx \int_0^{+\infty} x^{-2/\alpha} e^{-x} dx = \Gamma(1+2/\alpha)\Gamma(1-2/\alpha)$$
$$= 2/\alpha \Gamma(2/\alpha)\Gamma(1-2/\alpha) = [\sin c(2\pi/\alpha)]^{-1} > 1.$$

where the last two equalities follow from the well-known properties of Gamma Function:
$\Gamma(x+1) = x\Gamma(x), x > 0;\ \Gamma(x)\Gamma(1-x) = \pi/\sin \pi x, 0 < x < 1.$

This concludes the proof of (48).    □